\documentclass{iau} 
\usepackage{graphicx}
\usepackage{hyperref}

\title[Searching for X-ray Pulsations from Neutron Stars Using NICER] 
{Searching for X-ray Pulsations from Neutron Stars Using NICER}

\author[Paul S. Ray, Zaven Arzoumanian, \& Keith C. Gendreau, et al.]   
{Paul S. Ray$^1$
Zaven Arzoumanian$^2$, 
\and 
Keith C. Gendreau$^2$, \\
for the NICER Working Group on Pulsation Searches \\ and Multiwavelength Coordination}

\affiliation{$^1$U.S. Naval Research Laboratory, \\ Washington, DC 20375-5352\\ email: {\tt paul.ray@nrl.navy.mil} \\[\affilskip]
$^2$NASA/GSFC \\ Greenbelt, MD 20771}

\pubyear{2017}
\volume{337}  
\jname{Pulsar Astrophysics -­ The Next 50 Years}
\editors{P. Weltevrede, B.B.P. Perera, L. Levin Preston \& S. Sanidas, eds.}

\begin{document}

\maketitle

\begin{abstract}
The Neutron Star Interior Composition Explorer (NICER) presents an exciting new capability for 
exploring the modulation properties of X-ray emitting neutron stars, including large area, low 
background, extremely precise absolute event time stamps, superb low-energy response and flexible 
scheduling. The Pulsation Searches and Multiwavelength Coordination working group has designed 
a 2.5 Ms observing program to search for emission and characterize the modulation properties of 
about 30 known or suspected neutron star sources across a number of source categories. A key 
early goal will be to search for pulsations from millisecond pulsars that might exhibit thermal 
pulsations from the surface suitable for pulse profile modeling to constrain the neutron star equation 
of state. In addition, we will search for pulsations from transitional millisecond pulsars, isolated 
neutron stars, low-mass X-ray binaries (LMXBs), accretion-powered millisecond pulsars, central compact objects and other 
sources. We present our science plan and initial results from the first months of the NICER mission, including the discovery of pulsations from the millisecond pulsar J1231$-$1411.

\keywords{space vehicles: instruments, pulsars: general, (stars:) pulsars: individual (PSR J1231$-$1411), stars: neutron, X-rays: binaries}
\end{abstract}

\firstsection 

\section{Introduction}

While the majority of pulsar studies are done in the radio band, multiwavelength information is 
critical to gaining a fuller understanding of these objects.  Over the past decade this has been 
borne out by the spectacular success of the \textit{Fermi} mission and now about 10\% of the pulsar 
population is detected in gamma-rays, with a few dozen of those detections being radio-quiet
systems.  The Neutron Star Interior Composition Explorer (NICER) is poised to make similar
advances in studies of the X-ray emission from pulsars.

X-rays can provide a direct view of the surface of the star, giving information about the cooling
processes in young (or accretion-heated) neutron stars and the polar caps heated by magnetospheric return
currents in older pulsars. Pulse profiles from these sources encode information about the mass and radius
of the neutron star from special and general relativistic effects on the emitted radiation (see
Bogdanov et al., this volume). X-rays also probe non-thermal emission from the magnetosphere,
providing another handle on the system geometry and can reveal shock emission from interaction
of the pulsar wind with a companion star (see Roberts, this volume). Pulse timing studies
in the X-ray band are free from the propagation effects, such as time-variable interstellar dispersion and scattering,
that plague radio timing measurements.

In addition, some neutron stars are far more prolific X-ray emitters than radio. The largest classes
are the accreting systems like LMXBs and accreting pulsars, but this also includes most magnetars, 
isolated neutron stars and some radio quiet pulsars.

\section{NICER}

NICER is an X-ray telescope mounted on the International Space Station (ISS) that is highly optimized for
pulsar studies (\cite[Gendreau \etal\ 2016]{NICER}). NICER covers the 0.2 to 12 keV band with large collecting area ($>1900$ cm$^2$ at 
1.5 keV). This is twice the area of the XMM-Newton EPIC-pn camera that has been used for many 
pulsar studies. It detects and telemeters every event, with time stamps referenced to GPS to an 
accuracy of better than 100 ns.  The modular design with 56 co-aligned X-ray telescopes (52 currently functional on orbit)
results in very high count rate capability with low dead time. The inexpensive single-reflection
X-ray concentrator optics enable the use of small silicon drift detectors (SDDs), resulting in low radiation-induced 
background count rates. The SDDs also provide very good energy resolution of between 85 and 160 eV.
This combination of capabilities will be transformational for pulsar and neutron star studies
including pulsation searches, timing, spectral line studies, burst investigations and much more.

NICER was launched on 2017 June 3 and was robotically installed on the ISS ExPRESS Logistics Carrier (ELC) 2 on 2017 June 14. 
The first month was spent commissioning and calibrating the payload. NICER's 2-axis pointing system
slews at 1$^\circ$ per second and achieves an accuracy of 66 arcsec by means of a star camera co-aligned with
the X-ray telescope boresight. The individual telescopes are aligned to the overall boresight to an RMS of 28 arcsec. On 2017 July 17, NICER entered its science operations phase, which
is scheduled to last 18 months.  

The NICER Science Team is divided into several topical working groups whose purpose is to design a program
of observations that accomplishes a portion of the mission's overall science objectives (\cite[Arzoumanian \etal\ 2014]{NICERscience}). The working groups, with their chairpersons and time allocations are:
\begin{itemize}
\item Lightcurve Modeling --- Slavko Bogdanov (Columbia), 5.0 Ms
\item Bursts \& Accretion Phenomena --- Feryal Ozel (U. Arizona), 2.0 Ms
\item Searches \& Multiwavelength Coord. --- Paul Ray (NRL),  2.5 Ms
\item High-Precision Timing --- Andrea Lommen (Haverford), 4.0 Ms
\item Magnetars \& Magnetospheres --- Teru Enoto (Kyoto U.),  1.5 Ms
\item Observatory Science --- Ron Remillard (MIT),  2.5 Ms
\item Calibration --- Craig Markwardt (GSFC), 1.0 Ms
\end{itemize}

Coauthor KCG is NICER's Principal Investigator; ZA is Deputy PI and Science Lead. In this
paper, we will describe the work of the Searches  \& Multiwavelength Coordination working group.

\section{Searches  \& Multiwavelength Coordination}

The primary goal of this group is to achieve one of NICER's baseline science requirements, which
is to deeply search for pulsations from at least 20 pulsar candidates. We will achieve this
using coherent pulsation searches spanning all periods but targeting candidate MSPs especially.
The top priority for these searches is to discover X-ray pulsations from MSPs suitable for the
light curve modeling work to constrain the radius of a neutron star, but we have developed
a ranked list of sources that span over a dozen categories of candidate source classes.
Our final list includes 32 specific sources and reserves some time as well for transients and other sources
that may be identified during the mission.

\begin{figure}
 \includegraphics[width=\textwidth]{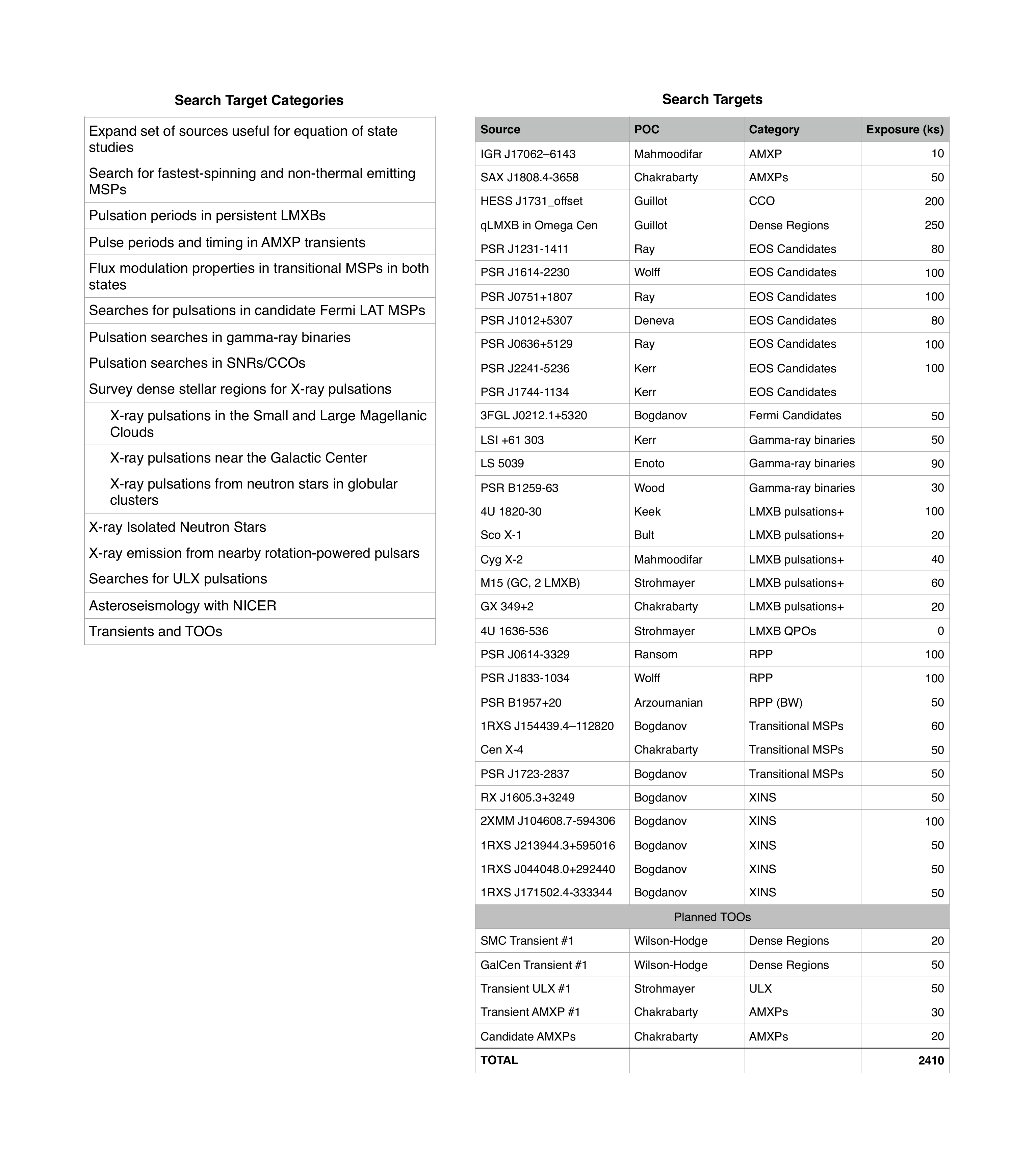}
 \caption{\textit{Left:} Source categories identified for the NICER Searching WG science program. \textit{Right:} Recommended target list for Searching WG time allocation.}
  \label{fig:sourcces}
\end{figure}

\section{Discovery of Pulsations from PSR J1231$-$1411}
 
The top priority source for this group is PSR J1231$-$1411. This 3.68 millisecond pulsar is one of the many radio MSPs discovered in searches of Fermi LAT unassociated sources. X-ray imaging observations with XMM-Newton revealed a moderately bright point source with a spectrum consistent with thermal emission from a neutron star surface (\cite[Ransom \etal\ 2011]{Ransom11}). Prior to NICER all X-ray observations of J1231$-$1411 were done with instruments/modes that did not have the timing resolution to determine if the emission was pulsed. A detection of pulsed emission from this source would make it a potentially important target for neutron star equation of state studies using pulse profile modeling. 
 
We analyzed data from 12 days of NICER observations of this source totaling a raw time of 127 ks. From that total exposure, we selected times where NICER was pointed at the target, which was at an elevation of $>30$ deg, and outside of the enhanced-background portions of the orbit (see Figure 2), resulting in 61.7 ks of exposure. We selected only photon events with energies $> 0.4$ keV and rejected likely particle energy-deposition events.

Pulse phases were computed using the \texttt{photonphase} code in PINT (\cite[Jing \etal\ 2017]{PINT}) and a radio timing ephemeris from \cite[Abdo \etal\ (2013)]{2PC}. We then found the energy cuts that optimized the H-test detection statistic (\cite[de Jager et al. 1989]{deJager89}). For an energy cut of 0.4--1.5 keV, we get an H-test value of 405, corresponding to an 18.6 $\sigma$ detection (not accounting for a small number of trials to optimize the energy cuts). Note that this is work in progress and will be revised as calibration and event filtering procedures are improved.

The pulsations are soft and highly sinusoidal, as expected from thermal emission from the surface, making PSR J1231$-$1411 a good addition to the targets for the NICER light curve working group.

\begin{figure}[b]
 \includegraphics[width=2.5in]{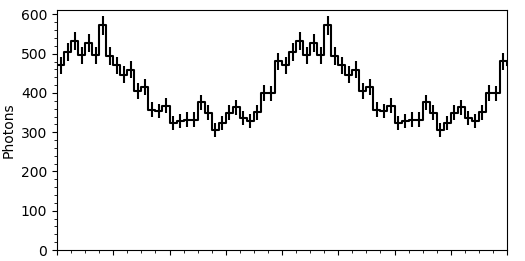} 
 \includegraphics[width=3.0in]{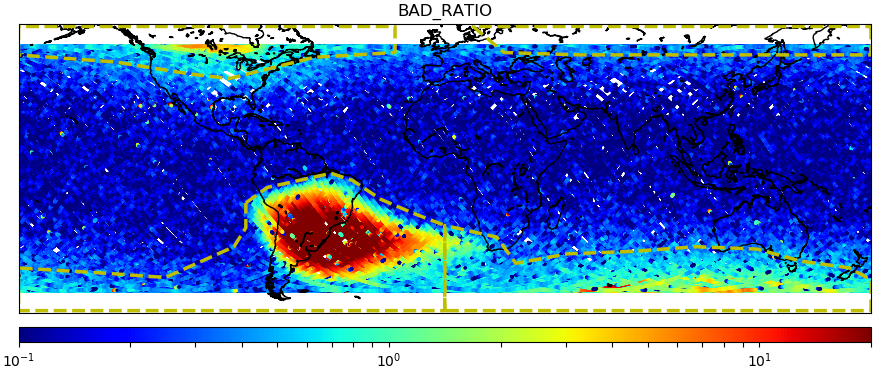} 
 \caption{\textit{Left:} Preliminary NICER 0.4--1.5 keV pulse profile (2 periods shown) of PSR J1231$-$1411.
 \textit{Right:} Map of a particle background proxy observed through the NICER orbit. The regions of the orbit inside the dashed polygons are excluded from the analysis to provide the lowest background rates.}
   \label{figlc}
\end{figure}

\section{Summary}

As the early discovery of PSR J1231$-$1411 demonstrates, NICER is now a powerful instrument
for X-ray studies of pulsars and other neutron star systems, and will be equally exciting for 
non-neutron star science such as AGN, black hole binaries, cluster spectroscopy, stellar coronae,
CVs and more. The community will have access to NICER through the Guest Observer program, 
which will issue its first call for proposals in early 2018. All NICER data will become public, with 
releases beginning in early 2018. Finally, the PI has significant discretionary time, so we
invite collaborations and requests for observations from the community. 

NICER is funded by NASA.

\end{document}